\documentclass[aps,prd,groupedaddress,nofootinbib,notitlepage,12pt]{revtex4}
\pdfoutput=1

\usepackage{graphicx}
\usepackage[mathscr]{euscript}
\usepackage{enumerate}
\usepackage{commands}

\begin{document}

\title{Bulk Locality and Entanglement Swapping in AdS/CFT}
\author{William R. Kelly}
\email{wkelly@ucdavis.edu}
\affiliation{Center for Quantum Mathematics and Physics (QMAP) \\
Department of Physics, University of California, Davis, CA 95616 USA}

\begin{abstract}
Localized bulk excitations in AdS/CFT are produced by operators which modify the pattern of entanglement in the boundary state.  We show that simple models---consisting of entanglement swapping operators acting on a qubit system or a free field theory---capture qualitative features of gravitational backreaction and reproduce predictions of the Ryu--Takayanagi formula.  These entanglement swapping operators naturally admit multiple representations associated with different degrees of freedom, thereby reproducing the code subspace structure emphasized by Almheiri, Dong, and Harlow.  We also show that the boundary Reeh--Schlieder theorem implies that equivalence of certain operators on a code subspace necessarily breaks down when non-perturbative effects are taken into account (as is expected based on bulk arguments).
\end{abstract}

\maketitle

\newpage 

\tableofcontents

\section{Introduction}

A central outstanding problem in holography is understanding how observables in ${\cal N}=4$ $SU(N)$ Super Yang--Mills map to observables in classical gravity in the low energy and large $N$ (i.e. semiclassical) limit.  Recently a great deal of attention has been paid to the fact that certain operators which are distinct at finite energies and finite $N$ become nearly indistinguishable in the semiclassical limit~\cite{Almheiri:2014lwa,Mintun:2015qda,Dong:2016eik,Bao:2016skw,Freivogel:2016zsb,Harlow:2016vwg}.  Careful study of these operators has lead to the construction of several qubit models of holographic states~\cite{Pastawski:2015qua,Yang:2015uoa,Hayden:2016cfa}.  These models also build off of the connection between tensor network states and the Ryu--Takayanagi formula originally put forward by Swingle~\cite{Swingle:2009bg,Swingle:2012wq}.

Like all models, the qubit models~\cite{Pastawski:2015qua,Yang:2015uoa,Hayden:2016cfa} capture some features of holography and fail to capture others.  For example, the HaPPY code~\cite{Pastawski:2015qua} contains a natural map between sets of ``bulk'' and ``boundary'' legs of a tensor network that is analogous to the map between bulk entanglement wedges and boundary regions in holography~\cite{Czech:2012bh,Wall:2012uf,Headrick:2014cta}.  The HaPPY code also contains local bulk operators associated with a particular bulk leg which can be ``pushed'' through the tensor network to any boundary region which contains the original leg in its entanglement wedge.  This latter feature does not exist in holography at large but finite $N$ and small but finite energy due to the gravitational dressing of bulk operators, as explained in~\cite{Almheiri:2014lwa} and reviewed in section~\ref{sec:review} below.

The purpose of this note is to present a simple model of holography which incorporates qualitative features of the gravitational dressing of bulk excitations.  In particular this means that different representations of a bulk operator must each be supported on the region where the gravitational dressing is anchored to the boundary (e.g. the region $B$ in Fig.~\ref{fig:dressing}).  Additionally, the focussing effects of gravitational flux and the Ryu--Takayanagi formula imply that the entanglement entropy of this overlap region must decrease.  We will show below that a model of holography where bulk excitations are created by swapping entanglement between boundary degrees of freedom naturally reproduces both of these features.  We also show that the inevitable breakdown of the code subspace picture due to non-perturbative effects is a natural consequence of the boundary Reeh--Schlieder theorem~\cite{Schlieder1965,haag1996local} in the continuum limit.

The prominent role that entanglement entropy plays in these models of holography is reminiscent of the connection between boundary entanglement and bulk dynamics developed in~\cite{Lashkari:2013koa,Faulkner:2013ica,Swingle:2014uza,Faulkner:2014jva,Kelly:2015mna,Lashkari:2015hha}.  We leave it to future work to fully implement these entanglement swapping operators in AdS/CFT and extract further lessons about the connection between entanglement and bulk physics.

\section{Bulk Excitations in Holography} \label{sec:review}

In this section we collect some well know facts about holography which will be important below.  In particular we review several features of perturbative bulk dynamics at large $N$.

\begin{figure}
\includegraphics[width=0.4 \textwidth]{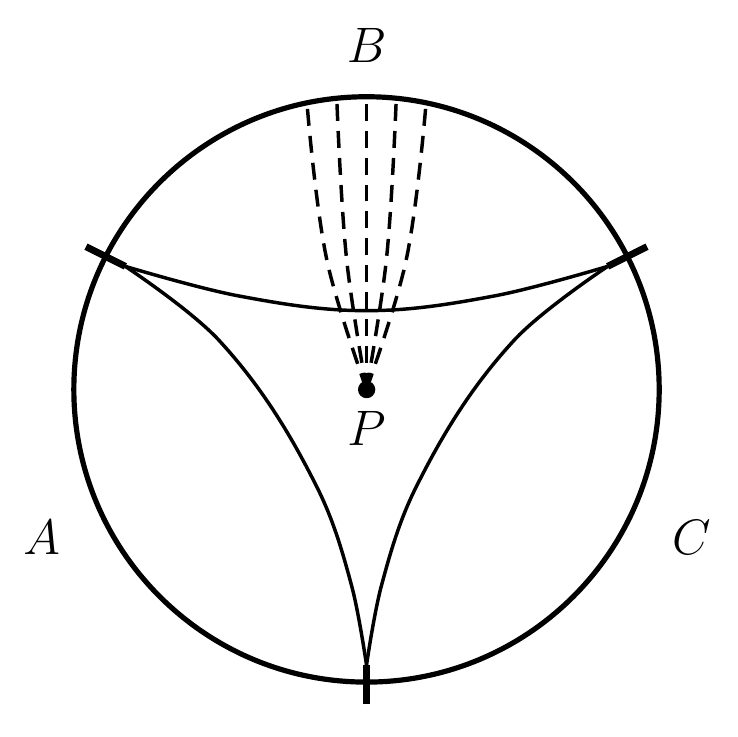} 
\caption{Schematic diagram of a bulk excitation localized around the point $P$ with gravitational dressing (dashed lines) focused into the boundary region $B$.  The entire bulk excitation---including gravitational dressing---extends beyond the entanglement wedge of $B$ but is contained in the entanglement wedges of $AB$ and $BC$.  The boundaries of the entanglement wedges are indicated with solid lines.}
\label{fig:dressing}
\end{figure}

To zeroth order in Newton's constant $G$ the bulk dynamics reduce to a non-gravitational field theory in which we can construct localized wave packets.  At leading order in $G$ these ``bare'' wave packets are ``dressed'' by the gravitational field.  As is always the case, the gravitational field contains both gauge and physical data.  Part of the physical data is the boundary stress tensor, which is determined by the asymptotic fall off of the field and is equal to the expectation value of the CFT stress tensor $\ev{T_{\mu\nu}}$.  For our purposes we will be interested in solutions for which $\ev{T_{\mu\nu}}=0$ outside of some ball shaped region $B$ (see Fig.~\ref{fig:dressing}).  When the bulk excitation extends beyond the entanglement wedge of $B$, as in Fig.~\ref{fig:dressing}, then the gravitational flux will reduce the area of the Ryu--Takayanagi surface.\footnote{I am not aware of a rigorous, non-perturbative proof of this claim in the classical bulk thoery, but it is true in simple cases and (as we will see in the next paragraph) it is a consequence of the Ryu--Takayanagi formula.}  Therefore, the bulk predicts that any operator which takes the vacuum to the state shown in Fig.~\ref{fig:dressing} must decrease the entanglement entropy $S_B$.

This story can be reproduced in the boundary theory using the HKLL construction~\cite{Balasubramanian:1998sn,Banks:1998dd,Balasubramanian:1998de,Bena:1999jv,Hamilton:2005ju,Hamilton:2006az,Kabat:2011rz,Heemskerk:2012mn,Heemskerk:2012np,Kabat:2012hp,Kabat:2012av,Kabat:2013wga,Morrison:2014jha}.  Certain smeared boundary operators produce localized bulk excitations, which by virtue of disturbing the vacuum, ensure that $\ev{T_{\mu\nu}}\ne 0$ somewhere on the boundary.  By adding additional smeared operators it is possible modify the gravitational dressing order by order in $1/N$~\cite{Kabat:2011rz,Heemskerk:2012mn,Kabat:2012av,Kabat:2013wga} to ensure that $\ev{T_{\mu\nu}}\ne 0$ only in region $B$.  It then follows from a straightforward calculation that acting on the vacuum with this operator reduces $S_B$, since
\begin{align} \label{eq:DeltaS}
\Delta S_B = \Delta S_{AC} = \Delta \ev{ H_{AC} } - S(\rho^{\vphantom{\text{vac}}}_{AC} | \rho_{AC}^\text{vac}) < 0 \ .
\end{align}
The first equality follows because both the initial and final states are pure and the second equality follows from the definition of the modular Hamiltonian $H$ and the relative entropy $S(\sigma|\rho)$
\begin{align}
H = - \log(\rho) \ , \qquad
S(\rho | \sigma ) =\tr\lb H (\rho-\sigma)\rb -  \lb S(\rho) - S(\sigma) \rb \ .
\end{align}
Since $AC$ is the complement of a ball shaped region, the modular Hamiltonian $ \Delta H_{AC}$ is a smeared integral of $\ev{T_{\mu\nu}}$ over $AC$~\cite{Bisognano:1975ih,Bisognano:1976za,Casini:2011kv}, which vanishes by construction.  Therefore the final inequality in~\eqref{eq:DeltaS} follows because $\Delta \ev{ H_{AC} }=0$ and the relative entropy is positive.  The inequality is strict because we have assumed $\rho^{\vphantom{\text{vac}}}_{AC} \ne \rho_{AC}^\text{vac}$, which implies $S(\rho^{\vphantom{\text{vac}}}_{AC} | \rho_{AC}^\text{vac}) > 0$.

It was argued in~\cite{Almheiri:2014lwa} that bulk configurations like that in Fig.~\ref{fig:dressing} have multiple boundary representations.  In particular such a representation exists on any boundary region which contains the entire bulk excitation (including gravitational dressing) in its entanglement wedge.  For example in the situation of Fig.~\ref{fig:dressing}, there exists unitary operators $U_\text{global},U_{AB},U_{BC}$ supported on the entire boundary, the region $AB$, and the region $BC$ respectively such that
\begin{align} \label{eq:EC}
U_\text{global}  \ket{\Omega} \sim U_{AB} \ket{\Omega} \sim U_{BC}\ket{\Omega}  \ ,
\end{align}
where $\ket{\Omega}$ is the vacuum state and here $\sim$ means ``equal to all orders in perturbation theory.''  This perturbative equality is expected to break down when non-perturbative effects are included, since the spacetime picture on which~\eqref{eq:EC} is based is no longer reliable.

One thing that~\eqref{eq:EC} makes clear is that the reduced density matrices $\rho_A$ and $\rho_C$ are not perturbatively affected by acting on the vacuum with any of any of the operators $U_\text{global} ,U_{AB} , U_{BC} $.  This is manifest because $\rho_A$ is not modified by $U_{BC} $, but $U_{AB} $ and $U_{BC}$ are perturbatively equivalent.  On the other hand, because the bulk excitation extends beyond the entanglement wedge of $B$ the operator $U_{AB}$ must somehow act non-trivially on $A$, even at finite order in perturbation theory.\footnote{otherwise we would have $\Delta S_B = 0$, contradicting~\eqref{eq:DeltaS}.}  The central insight of this paper is that $U_{AB}$ acts on $A$ by modifying how the degrees of freedom in $A$ are entangled with the rest of the state.  In section~\ref{sec:Eswap} we will construct  models that realize this idea.

\section{Quantum Effects}

In the previous section we argued that the boundary result~\eqref{eq:DeltaS} is consistent with the classical bulk theory.  Since~\eqref{eq:DeltaS} is an exact result it should also hold in the bulk when we include quantum effects.  In the semi-classical limit the entanglement entropy can be expanded in the form
\begin{gather}
\Delta S = \Delta S_\text{classical} + \Delta S_\text{one-loop} + \dots  \ ,
\end{gather}
where the dots indicate higher loop corrections.  $\Delta S_\text{one-loop}$ was compute in~\cite{Faulkner:2013ana} and shown to be
\begin{gather}
\Delta S_\text{one-loop} = \frac{ \Delta \langle \hat A \rangle }{4 G} + \Delta S_\text{bulk} \ ,
\end{gather}
where $ \Delta S_\text{bulk}$ is the change in the entanglement entropy of the associated bulk entanglement wedge and $\hat A$ is a linear operator with expectation value equal to the area of the classical Ryu--Takayanagi surface.\footnote{In certain cases $\langle \hat A \rangle$ includes additional ``Wald-like'' terms, see~\cite{Faulkner:2013ana} for details.  Here we focus on the case where $\langle \hat A \rangle$ gives the area.}

Now consider a quantum perturbation to the vacuum state of the type depicted in Fig.~\ref{fig:dressing}.  For such a perturbation $(\Delta S_B)_\text{classical} = 0$ and therefore~\eqref{eq:DeltaS} implies that
\begin{align} \label{eq:QBB}
\Delta S_\text{bulk} < -\frac{ \Delta \langle \hat A \rangle }{4 G} \ ,
\end{align}
for the bulk entanglement wedge and Ryu--Takayanagi surface associated with the boundary region $B$.  As luck would have it~\eqref{eq:QBB} follows immediately from the Quantum Bousso bound derived in~\cite{Bousso:2014sda}.  The Quantum Bousso bound applies here because $\Delta S_\text{bulk}$ can be computed on any Cauchy surface of the bulk entanglement wedge, including the past light sheet emanating from the Ryu--Takayanagi surface.  Evidently the bulk semiclassical theory knows (at least at one-loop order) that it must obey~\eqref{eq:DeltaS}.\footnote{I thank an anonymous referee for suggesting that I consider quantum effects.}

\section{Entanglement Swapping} \label{sec:Eswap}

We now construct two simple models that reproduce the features of holography described in section~\ref{sec:review}.  The first model is a qubit model and the second is a free field theory model which in principle could be adapted to apply to real holographic systems, though the details will not be worked out here.  Both models will rely on entanglement swapping operator that leave the reduced density matrix of a particular subsystem unchanged.

\subsection{Qubit Model}

\begin{figure}
\includegraphics[width=0.8 \textwidth]{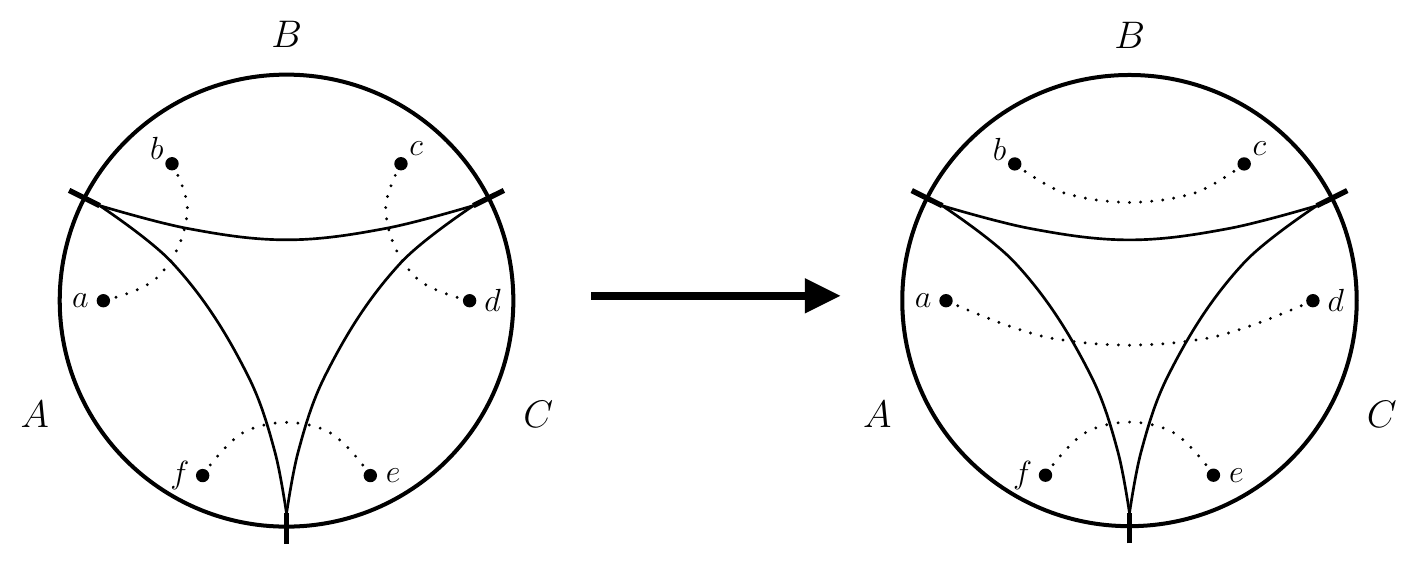} 
\caption{(left) Diagram of the qubit ``vacuum state''~\eqref{eq:qubitVacuum} with dotted lines signifying Bell paris. (right) The resulting state after any of the three unitary operators~\eqref{eq:qubitOps} act on the vacuum state.  Note that the resulting state can be obtained by swapping qubits $a$ and $c$ or $b$ and $d$.  In this diagram the entanglement entropy of any region is equal to $\log(2)$ times the number of dotted lines crossing the boundary of the associated entanglement wedge.}
\label{fig:swap}
\end{figure}

Consider the six qubit system depicted in Fig.~\ref{fig:swap}.  The subsystems $A$, $B$, and $C$ represent spatial regions in the boundary theory and the dashed lines signify Bell pairs, indicating that the initial state of the system is
\begin{align} \label{eq:qubitVacuum}
\ket{\Omega} := \lp \ket{0_a 0_b} + \ket{1_a 1_b } \rp \otimes \lp \ket{0_c 0_d} + \ket{1_c 1_d } \rp \otimes \lp \ket{0_e 0_f} + \ket{1_e 1_f } \rp  \ .
\end{align}
This state will represent the vacuum state of the boundary CFT.

In this model, we represent the operator $U_\text{global}$, $U_{AB}$, and $U_{BC}$ from~\eqref{eq:EC} with the operators
\begin{align} \label{eq:qubitOps}
U^\text{qubit}_\text{global} &:= {\bigg [} \ket{1001}\bra{0011}+\ket{0110}\bra{1100} + (\text{h.c.}) {\bigg ]}_{abcd} \otimes \mathbb{I}_{ef} + \mathbb{I}_\perp \cr
U^\text{qubit}_{AB} &:= {\bigg [} \lp \ket{100}\bra{001}  + \ket{011}\bra{110} \rp \otimes \mathbb{I}_d  + (\text{h.c.}) {\bigg ]}_{abcd} \otimes \mathbb{I}_{ef} + \mathbb{I}_\perp \cr
U^\text{qubit}_{BC} &:= {\bigg [}  \mathbb{I}_a  \otimes \lp \ket{110} \bra{011} + \ket{110}\bra{100} \rp  + (\text{h.c.}) {\bigg ]}_{abcd} \otimes \mathbb{I}_{ef} + \mathbb{I}_\perp \ ,
\end{align}
where in each line $\mathbb{I}_\perp$ is the identity operator on the kernel of the first term and $(\text{h.c.})$ denotes the Hermitian conjugate.  The choice of $\mathbb{I}_\perp$ is arbitrary and it could be replaced by any unitary $U_\perp$ (and $\mathbb{I}_{ef}$ could similarly be replaced with some $U_{ef}$).

Right away it is trivial to verify that all three operators are unitary, inequivalent, and yet produce the same state when acting on the vacuum.
We can easily see that all three operators are inequivalent by evaluating specific matrix elements, for example
\begin{align}
\Braket{100\alpha}{ U^\text{qubit}_\text{global} }{001\alpha} &= \delta_{\alpha,1} \mathbb{I}_{ef} \cr
\Braket{100\alpha}{ U^\text{qubit}_{AB} }{001\alpha} &= \mathbb{I}_{ef} \cr
\Braket{100\alpha}{ U^\text{qubit}_{BC} }{001\alpha} &= 0  \ ,
\end{align}
and a simple calculation gives
\begin{align} \label{eq:ECqubit}
&U^\text{qubit}_\text{global} \ket{\Omega} = U^\text{qubit}_{AB} \ket{\Omega} = U^\text{qubit}_{BC} \ket{\Omega} 
\cr & \qquad =  \lp \ket{0_a 0_d} + \ket{1_a 1_d } \rp  \otimes \lp \ket{0_b 0_c} + \ket{1_b 1_c } \rp \otimes \lp \ket{0_e 0_f} + \ket{1_e 1_f } \rp   \ .
\end{align}
Examining Fig.~\ref{fig:swap}, it is clear that swapping qubits $a\leftrightarrow c$ or $b\leftrightarrow d$ results in exactly the same final state, and this is the reason why the final state can be created by acting on $AB$ or $BC$.  For this reason we will refer to the operators~\eqref{eq:qubitOps} as entanglement swapping operators.

The exact equality in~\eqref{eq:ECqubit} is stricter than the perturbative equality we required in~\eqref{eq:EC}.  We will see below that this discrepancy is naturally taken care of when we pass to the continuum limit.

Finally while the reduced density matrices of $\rho_A$ and $\rho_C$ are unchanged by the operators~\eqref{eq:qubitOps}, $\rho_B$ is clearly altered.  In fact the subsystem $B$ is now in a pure state, which means that
\begin{align}
\Delta S_B = -2 \log 2 \ ,
\end{align}
reproducing the fact that $\Delta S_B < 0$ derived in section~\ref{sec:review}.  To make the model slightly more realistic we could replace each Bell pair in~\eqref{eq:qubitVacuum} with $O(N^2)$ Bell pairs and have the entanglement operators act on a single pair of qubits as in~\eqref{eq:qubitOps}.  In this model we would have $S_B\sim N^2$ and $\Delta S_B \sim N^0$, as expected when we add a single bulk particle to vacuum of AdS/CFT.

The above qubit model is reminiscent of the qutrit model of local bulk operators in~\cite{Almheiri:2014lwa}, however there are several qualitative differences.  Most importantly, the operators~\eqref{eq:qubitOps} have a common overlap region $B$ and it is impossible to create the final state $U^\text{qubit}_\text{global} \ket{\Omega}$ without acting on $B$.  This is in contrast to the qutrit model of~\cite{Almheiri:2014lwa} where any excited state can be created by acting on any two regions, including $AC$.  A second (related) difference is that in the qutrit model there are no single subsystem measurements which indicate that the state has changed, one must perform a joint measurement on at least two regions.  Here, the excitation can be detected (though not fully reconstructed) by making local measurements on $B$.  These local measurements are analogous to measuring $\ev{T_{\mu\nu}} $ in the boundary theory, and it is necessary that some measurement in $B$ distinguish the final state from the vacuum in order to have $\Delta S_B < 0$.

\subsection{Free Field Theory Model}

Consider a free CFT on a sphere.  Let $B$ be a ball centered on the north pole.  The vacuum state can be be decomposed with respect modes on the region $B$ and the complementary region $\bar B$ in the form
\begin{align} \label{eq:Rindler}
\ket{\Omega} 
&= Z^{-1} \bigotimes_k \lp \sum_{n=0}^\infty e^{-E_{n,k}/2} \ket[k]{ n_{\vphantom{\bar B}B} , n_{\bar B} } \rp  \ ,
\end{align}
where $k$ runs the eigenmodes of the ``boost'' Hamiltonian, $n_{\vphantom{\bar B}B} , n_{\bar B}$ are the mode occupation numbers, $E_{n,k}$ are the associated energies, and $Z$ is a normalization constant~\cite{Unruh:1976db}.

To recreate the essential features of the qubit model, it will be useful to pick out two modes $k$ and $l$ and to identify the four associated occupation numbers with the labels $a,b,c,d$ as follows
\begin{align}
\ket{\Omega} = Z^{-1} \lp \sum_{n} e^{-E_{n,k}/2} \ket[k]{n_a,n_b} \rp \otimes \lp \sum_{m} e^{-E_{m,l}/2} \ket[l]{m_c,m_d} \rp \otimes \dots \ ,
\end{align}
where, as in Fig.~\ref{fig:swap}, $b$ and $c$ are localized in subsystem $B$ while $a$ and $d$ are not.  Unlike in Fig.~\ref{fig:swap}, the modes $a$ and $d$ are not localized into separate spatial regions $A$ and $C$, but rather are each supported on the whole of $AC$.

Now consider the three operators
\begin{align} \label{eq:fftOps}
U^\text{f.f.}_\text{global} &:= \lb \sum_{n\ne m} \ket{m,n,n,m}\bra{n,n,m,m} + (\text{h.c.})\rb_{abcd} \otimes \mathbb{I}_\text{rest} + \mathbb{I}_\perp \cr 
U^\text{f.f.}_{AB} &:=  \lb \sum_{n\ne m} \ket{m,n,n}\bra{n,n,m} \otimes \mathbb{I}_d + (\text{h.c.})\rb_{abcd}  \otimes \mathbb{I}_\text{rest} + \mathbb{I}_\perp \cr
U^\text{f.f.}_{BC} &:=  \lb \sum_{n\ne m} \mathbb{I}_a \otimes \ket{m,m,n}\bra{n,m,m}   + (\text{h.c.})\rb_{abcd} \otimes \mathbb{I}_\text{rest}  + \mathbb{I}_\perp 
\end{align}
where $\mathbb{I}_\text{rest}$ is the identity operator on all degrees of freedom other than $a,b,c,d$ and, as before, $\mathbb{I}_\perp$ is defined in each line to be the identity operator on the kernel of the first term.  For now the labels $AB$ and $BC$ are aspirational and don't refer to any particular spatial region.

As before, it is trivial to verify that the operators are unitary, inequivalent, and yet produce the same state when acting on the vacuum.  We can see that they are inequivalent by considering the matrix elements
\begin{align}
\Braket{m,n,n,p}{ U^\text{f.f.}_\text{global} }{n,n,m,p} &= \delta_{p,m} \mathbb{I}_{rest} \cr
\Braket{m,n,n,p}{ U^\text{f.f.}_{AB} }{n,n,m,p} &= \mathbb{I}_{rest} \cr
\Braket{m,n,n,p}{ U^\text{f.f.}_{BC} }{n,n,m,p} &= 0 \ ,
\end{align}
for $p\ne n$ and $n \ne m$, and a simple calculation gives
\begin{align} \label{eq:ECff}
&U^\text{f.f.}_\text{global} \ket{\Omega} = U^\text{f.f.}_{AB} \ket{\Omega} =U^\text{f.f.}_{BC} \ket{\Omega} 
\cr & \qquad = Z^{-1} \lp \sum_{n,m} e^{-(E_{n,k}+E_{m,l})/2} \ket[k]{m_a,n_b} \otimes \ket[l]{n_c,m_d} \rp \otimes \dots \ .
\end{align}
As before it is straightforward to verify that $\rho_A$ and $\rho_C$ (defined for now by tracing out all degrees of freedom except $a$ and $d$ respectively) are unchanged by the operators~\eqref{eq:fftOps}.  On the other hand $\rho_B$ (defined by tracing out all degrees of freedom except $b$ and $c$) has been modified and the entanglement entropy $S_B$ has decreased.  This follows from observing that the modes $b$ and $c$ are now in a pure state, whereas before they were completely uncorrelated.

As in the qubit model,~\eqref{eq:ECff} is exact even though we only needed perturbative equality, however this time we have an explanation.  For simplicity we have so far worked with eigenstates of the modular Hamiltonian, which allowed us to write $\ket{\Omega}$ in the simple form~\eqref{eq:Rindler}.  For that reason the modes $a$ and $d$ are not restricted to any particular spatial region within $AC$.  However we could have chosen to work with a spatially localized set of modes restricted to non-overlapping spatial regions.  This would have made~\eqref{eq:Rindler} more complicated, but the modes would still be entangled and entanglement swap operators could still be constructed.  However, as long as the combined spatial support of $U_{AB}$ and $U_{BC}$ is not the entire boundary---as would be the case if both operators were supported in the interior of their respective spatial regions---then a corollary to Reeh--Schlieder theorem (theorem 5.3.2 of~\cite{haag1996local}) states that
\begin{align}
(U_{AB} - U_{BC} ) \ket{\Omega} \ne 0 \ .
\end{align}
That means the best we can possibly do is
\begin{align}
U_{AB} \ket{\Omega} \sim U_{BC} \ket{\Omega}  \ .
\end{align}
Thus the non-perturbative breakdown mentioned above is a generic feature of continuum quantum field theories and does not require any special properties of the field theory or operators.  It would be interesting to try to formulate an approximate Reeh--Schlieder theorem and place a lower bound on the size of non-perturbative effects.

\section{Discussion}

Modeling of qasilocal bulk operators in AdS/CFT as entanglement swapping operators in the boundary theory provides a simple framework that ties together the existence of multiple boundary representations of a single bulk operator, basic features of gravitational backreaction, and the Ryu--Takayanagi formula.  Throughout we have focused on perturbations about the vacuum state primarily because special properties of the vacuum state allow us to derive the useful inequality~\eqref{eq:DeltaS} and the decomposition~\eqref{eq:Rindler} (which both are due to the simple form of the modular Hamiltonian of $\rho_B$).  Obviously it would be desirable to model non-vacuum states, and there is no obvious obstruction to doing so.  However without~\eqref{eq:DeltaS} and~\eqref{eq:Rindler} such models are even more schematic and harder to verify even qualitatively, thus we leave thinking about non-AdS spacetimes for future work.

Still the question remains, can the HKLL operators in AdS/CFT actually be understood as swapping entanglement between different spatial regions of the boundary?  On some level the answer must be yes by the argument given in the last paragraph of section~\ref{sec:review} above.  On the other hand, it would be valuable to understand how this works in detail in a strongly interacting, large $N$ CFT.  One obstacle is that it is difficult to explicitly write down HKLL operators with tightly collimated gravitational dressing (as in Fig.~\ref{fig:dressing}), though it may be possible to make progress in $\text{AdS}_3/\text{CFT}_2$ since the gravitational field has no propagating degrees of freedom (see~\cite{Donnelly:2015taa}).

The goal of this program would be to develop a non-perturbative framework for thinking about quasilocal bulk operators.  This would be valuable because, while the connection between entanglement and bulk dynamics is well understood to leading order in perturbation theory~\cite{Lashkari:2013koa,Faulkner:2013ica}, new tools are needed to understand non-linear, classical gravity in the bulk.

\section*{Acknowledgements}

It is a pleasure to thank Ahmed Almheiri, Ning Bao,
Xi Dong, 
Netta Engelhardt, 
Felix Haehl, 
Daniel Harlow, 
Gavin Hartnett, Veronika Hubeny, Don Marolf, 
Jonathan Oppenheim,
Mukund Rangamani, and Max Rota for helpful discussions and feedback.  
I am also grateful to the organizers and participants of the ``Quantum Information in String Theory and Many-body Systems'' workshop at the Yukawa Institute for Theoretical Physics, where these ideas were partially developed.
This work was supported by funds from the University of California.

\bibliographystyle{kp}

\bibliography{EntanglementSwapV6.bbl}

\end{document}